\documentclass[aps,prd,preprintnumbers,twocolumn]{revtex4-1}
\usepackage{epsfig}
\usepackage{latexsym,amsmath,amssymb,amstext}
\usepackage{multirow}

\usepackage{xcolor}
\usepackage{soul}
\definecolor{hu-berlin-blue}{RGB}{0,65,137} 
\newcommand{\jhw}[1]{\textcolor{hu-berlin-blue}{#1}} 

\begin{document}
\title{The bottomonium melting from screening correlators at high temperature}
\preprint{HU-EP-21/23-RTG}
\author{Peter Petreczky}
\affiliation{Physics Department, Brookhaven National Laboratory, Upton NY 11973}
\author{Sayantan Sharma}
\affiliation{The Institute of Mathematical Sciences, HBNI, Chennai, 600113}
\author{Johannes Heinrich Weber}
\affiliation{Institut f\"ur Physik, Humboldt-Universit\"at zu Berlin \& IRIS Adlershof, D-12489 Berlin, Germany}
\begin{abstract}
We study the bottomonium screening masses in a 2+1 flavor QCD on the lattice using the Highly 
Improved Staggered Quark(HISQ) action.  We focus on a wide temperature range
in the region $350~{\rm MeV}\le T < 1000 ~{\rm MeV}$, and perform our calculations on three different 
lattice spacings corresponding to temporal lattice extent of 
$N_\tau=8,10$ and $12$, in order to control the lattice cut-off effects.
From a detailed study of the temperature dependence of screening masses we conclude that 
the $\eta_b(1S)$ and $\Upsilon(1S)$ states melt at $T>500$ MeV, while the scalar and axial-vector states 
$\chi_{b0}(1P)$ and $h_b(1P)$ melt already at $T>350$ MeV.
\end{abstract}

\maketitle

\section{Introduction}

The study of the properties of bound states of heavy quarks called quarkonium  
has received a lot of attention since the seminal paper by Matusi and Satz~\cite{Matsui:1986dk}, 
where it was suggested that the formation of a color deconfined medium in heavy-ion collisions
will lead to the dissolution of quarkonium, resulting in a suppression of their production. There are large 
experimental efforts in recent times dedicated towards understanding  quarkonium production 
and its dynamics in heavy-ion collisions, which are supplemented by many phenomenological 
studies, see Refs.~\cite{Sharma:2021vvu,Yao:2021lus,Rothkopf:2019ipj,Aarts:2016hap,Mocsy:2013syh,Bazavov:2009us} for recent reviews.

The in-medium properties of quarkonium in a Quark-Gluon plasma (QGP) as well as its 
dissolution as a function of the temperature are all encoded in the quarkonium spectral functions,
that are defined in terms of the real-time correlation functions of the appropriate hadron 
operators. Quarkonium states show up as peaks in the spectral functions, which at 
high temperatures are expected to be broadened and shifted in the frequency space, and 
eventually merge into the continuum of quark-antiquark scattering states. Through an analytic 
continuation one can relate the spectral function of the quarkonium state  of a specific quantum 
number channel to the Euclidean time correlation function, which can be calculated using lattice
field theory techniques. The Euclidean time correlation function is the Laplace transform of the 
spectral function, if thermal particle production can be neglected,  and has a more complicated 
kernel otherwise. Therefore, lattice calculations can in principle provide a model independent 
information on the quarkonium spectral functions. However, in practice the reconstruction of the 
spectral function from a discrete set of Euclidean correlator data from lattice is a difficult task. 
Early works on the reconstruction of the spectral functions have been reported in Refs. 
~\cite{Nakahara:1999vy,Asakawa:2000tr,Asakawa:2003re,Wetzorke:2001dk,Karsch:2002wv,
Umeda:2002vr,Datta:2003ww,Jakovac:2006sf}.  It has also been pointed out that the Euclidean 
correlation functions have limited sensitivity to the in-medium quarkonium 
properties and/or their melting~\cite{Mocsy:2007yj,Petreczky:2008px}, because at high temperatures the temporal extent
of the lattice becomes small. In the case of bound states of bottom quarks (bottomonium) there 
is an additional problem of large discretization errors in the correlators due to the large bottom-quark 
mass. One can circumvent this problem by using non-relativistic QCD (NRQCD), an effective theory in 
which the energy scale associated with the heavy quark mass has been integrated out \cite{Caswell:1985ui}.
This approach is widely used to calculate bottomonium 
properties at zero temperature ~\cite{Lepage:1992tx,Davies:1994mp,Meinel:2009rd,Meinel:2010pv,Hammant:2011bt,Dowdall:2011wh}.
Recent studies within the NRQCD formalism at finite temperature 
~\cite{Aarts:2010ek,Aarts:2014cda,Kim:2014iga,Kim:2018yhk,Larsen:2019bwy,Larsen:2019zqv}
have indicated that the ground states of bottomonium channels, $\Upsilon(1S)$ and $\eta_b(1S)$ can survive
up to temperatures of $400$ MeV, whereas the fate of $P$-wave bottomonia is not 
yet completely settled. Lattice results from the FASTUM collaboration suggest that the 
$P$-wave states melt  already at temperatures around $200$ MeV \cite{Aarts:2014cda}, while other independent 
studies also within the NRQCD formalism suggest that $P$-wave bottomonia 
can survive at higher temperatures within the QGP \cite{Kim:2018yhk,Larsen:2019bwy,Larsen:2019zqv}.
Since NRQCD is an effective theory with an ultra-violet cutoff of around the bottom quark mass, 
the choice of lattice spacings in these calculations cannot be too small. Since the temperature is related 
to the inverse lattice spacing, $T=1/(a N_{\tau})$ with $N_{\tau}$ being the temporal extent, this also  
suggests that studying bottomonia at high temperature with some reasonable choice of $N_\tau$ is 
difficult. For this reason  we do not know yet at what temperatures the ground state bottomonia melt.

The spatial correlation functions of mesons can offer a different perspective on the problem 
of in-medium modification of mesons, in particular 
charmonium~\cite{Karsch:2012na,Bazavov:2014cta,Bazavov:2020teh}. In contrast to the temporal 
meson correlators, the spatial meson correlation functions can be calculated for large (spatial) 
separations between the source and the sink and, therefore, are more sensitive to the in-medium
modifications of meson states~\cite{Karsch:2012na,Bazavov:2014cta,Bazavov:2020teh}.  The 
spatial correlation functions are in turn related to the meson spectral function at non-zero 
momenta through the relation
\begin{equation}
G(z,T)=\int_{0}^{\infty} \frac{2 d \omega}{\omega} \int_{-\infty}^{\infty} d p_z e^{i p_z z} \sigma(\omega,p_z,T).
\label{eq:spatial}
\end{equation}
While the above relation is more complicated than the corresponding relation for the temporal correlator 
and spectral function for mesons,  it is still quite useful. At large distances the spatial meson correlation 
function decays exponentially, and the exponential decay is governed by the so-called screening mass, 
$G(z) \sim \rm{exp}(-M_{scr} z)$. When there is a well-defined bound state peak in the meson spectral function,  
the screening mass will be equal to the meson pole mass \cite{Karsch:2012na,Bazavov:2014cta}. On the other hand at very high temperatures, 
when the quark and antiquarks are eventually unbound, the screening mass is given by 
$2\sqrt{(\pi T)^2+m_q^2}$, with $m_q$ being the quark mass. Thus the temperature 
dependence of the meson screening masses can provide some valuable information about the melting of 
meson states. The analysis of the spatial correlation functions have provided some evidence for sequential 
in-medium modification of different charmonium states, i.e. stronger in-medium modification of excited 
charmonia compared to its ground states, and for the dissolution of the $1S$ charmonium state 
at temperatures $T>300$ MeV~\cite{Bazavov:2014cta}.

The aim of this paper is to provide some new insights on the melting of bottomonium states in the QGP through 
the study of their spatial correlation functions. For the first time we use the full relativistic Dirac operator 
for the bottom quarks in the construction of the meson correlators in 2+1 flavor QCD, which allows us to make an independent prediction 
on the melting of different quantum number states independent of the NRQCD formalism.  We can thus unambiguously 
observe an earlier melting of the scalar and axial-vector bottomonium states compared to the pseudo-scalar and 
vector channels. The  paper is organized as follows. In section II we provide the details of the techniques we use.  
Subsequently the main results on the bottomonium screening masses are discussed in section III, followed by our 
concluding section. 

\section{Lattice setup}
We calculate the screening masses of the bottomonium states in QCD with 
$2+1$ flavors of dynamical quarks treating the bottom quarks in the quenched  
approximation.  We use the Highly Improved Staggered Quark (HISQ) action~\cite{Follana:2006rc} 
for the quarks and a tree level Symanzik improved gauge action.  Using HISQ action for the 
valence bottom quark is important since it preserves the correct dispersion relation for heavy quarks 
~\cite{Follana:2006rc}. The strange quark mass, $m_s$ was chosen to be close to its physical value, 
while the light quark masses $m_l=m_s/20$ correspond to a pion mass of $160$ MeV in 
the continuum limit~\cite{Bazavov:2014pvz}. We perform our calculations on $N_{\sigma}^3\times N_{\tau}$ 
lattice with temporal extent of $N_\tau=8,10,12$ and the spatial extent, $N_{\sigma}$ fixed by $N_{\sigma}=4 N_\tau$.
The corresponding gauge configurations have been generated by the HotQCD collaboration~\cite{Bazavov:2013uja,Bazavov:2014pvz,Ding:2015fca}. We have specifically focused 
on a wide temperature range between $2-8~T_c$, where $T_c=156.5(1.5)$ MeV is the chiral crossover 
temperature~\cite{HotQCD:2018pds}. This enables us to measure the full details of the thermal evolution 
of the bottomonium correlators. Moreover we ensured that $m_b a\lesssim 1$ for the lattice spacings 
over this entire temperature range of interest, which in turn allowed us to have sufficient control on the lattice artifacts 
in the results of the bottomonium correlators. Having three different lattice extents allowed us to  
have a better control on the discretization effects at high temperatures. The bottom-quark mass in this 
entire range was set to be $52.5m_s$, which is close to its physical value. The lattice spacing was determined 
in physical units using the $r_1$ scale defined in terms of the static quark\jhw{-}antiquark potential through
$\left . r^2 \frac{dV(r)}{dr}\right|_{r=r_1}=1$. 
We used the parametrization of $a/r_1$ obtained in Ref. ~\cite{Bazavov:2017dsy} and the value 
$r_1=0.3106(18)$ fm~\cite{MILC:2010hzw}. The details of the lattice parameters including the 
bare lattice gauge coupling $\beta=10/g_0^2$, the quark masses, the temperatures as well as the number of 
configurations used in this work are summarized in Table~\ref{tab:latpar}.

\begin{center}
\begin{table}[h]
\begin{tabular}{|l|l|l|l|l|l|l|l|l|}
\hline
$\beta$ & $a m_s$ & $am_b$ & \multicolumn{2}{|c|}{$N_\tau=8$}  & \multicolumn{2}{c|}{$N_\tau=10$} & \multicolumn{2}{c|}{$N_\tau=12$} \\ 
\hline
      &       &        & $T$    &\# c    & $T$       &\# c       & $T$       &  \# c     \\
\hline 
7.650 & 0.0192& 1.0090 & -      & -      & -         &  -        & 350       &   220	  \\	
\hline
7.825 & 0.0164& 0.8618 & 611    & 500    & 489       & 250	  & 407       &   180	  \\	
\hline
8.000 & 0.0140& 0.7357 & 711    & 500    & 571       & 500	  & 476	      &   180	\\	
\hline
8.200 & 0.0117& 0.6133 & 844    & 250    & 675       & 250	  & 562       &   500	\\	
\hline
8.400 & 0.0098& 0.5124 & 1000   & 240   & 800       & 250	  & 666       &   500	\\	
\hline
8.570 & 0.0084& 0.4402 & -      & -     & 923       & 250	  & 769       &   250	\\	
\hline
8.710 & 0.0074& 0.3886 & -      & -     & -         & -         & 866       &   250	\\	
\hline
8.850 & 0.0065& 0.3431 & -      & -	& -         & -         & 974       &   250	\\	
\hline
\end{tabular} 
    \caption{The gauge coupling, $\beta$, the quark masses, the temperature values and the number of 
gauge configurations (\#c) used in this study.}
    \label{tab:latpar}
\end{table}
\end{center}

The meson operators for staggered fermions have the form 
\begin{equation}
J_M(\mathbf{x})=\bar q(\mathbf{x}) (\Gamma_D \times \Gamma_F) q(\mathbf{x}), ~\mathbf{x}=(x,y,z,\tau),
\end{equation}
where $\Gamma_D, \Gamma_F$ are the Dirac gamma matrices corresponding to the spin and the 
staggered taste (flavor) structure. In this work we consider the case where $\Gamma_D=\Gamma_F=\Gamma$. 
This choice corresponds to local operators for the meson currents, which in terms of the staggered quark fields 
have the simple form $J_M(\mathbf{x})=\tilde \phi(\mathbf{x}) \bar \chi(\mathbf{x}) \chi(\mathbf{x})$. 
The staggered phase $\tilde \phi(\mathbf{x})$ specifies the quantum numbers of the meson channel.
In this work we consider the spatial correlation function\jhw{s} along the $z$-direction:
\begin{equation}
C_M(z)=\int dx dy d\tau \langle J_M(\mathbf{x}) J_M(0)\rangle .
\end{equation}
For each choice of $\tilde \phi(\mathbf{x})$, the staggered meson correlation function
contains contributions from both parity states, which correspond to the 
oscillating and non-oscillating parts of the correlators. If we restrict 
ourselves to the lowest states, the spatial meson correlation
function can be simply written as
\begin{eqnarray}
\nonumber
 C_M(z)&=&A_{NO} \cosh\left[M_{NO}\left(z-\frac{N_s}{2}\right)\right]\\
&-&(-1)^z 
 A_{O} \cosh\left[M_O\left(z-\frac{N_s}{2}\right)\right].
\label{fit}
\end{eqnarray}
The subscripts $O$ and $NO$ for the screening masses and amplitudes denote the oscillating
and non-oscillating states.
In Table~\ref{tab:meson} we give the details of the staggered phases corresponding to the oscillating 
and non-oscillating contributions for different meson quantum numbers, as well as the labels denoting the 
screening masses in the pseudo-scalar (PS), scalar (S), vector (V) and the axial-vector (AV) channels. 
\begin{table}
\begin{tabular}{cccccc}
\hline
                 &  $- \tilde \phi(\mathbf{x})$                       & $\Gamma$            & $J^{PC}$  & meson   & screening mass  \\
\hline
$M_{NO}$  & \multirow{2}{*}{$1$}                     & $\gamma_4 \gamma_5$ & $0^{-+}$  & $\eta_b$     & $M_{scr}^{PS}$ \\
$M_{O}$  &                                           & $ 1 $               & $0^{++}$  & $\chi_{b0}$  & $M_{scr}^{S}$ \\
\hline
$M_{NO}$  & \multirow{2}{*}{$(-1)^{x+\tau},~(-1)^{y+\tau}$} & $\gamma_i$    & $1^{--}$  & $\Upsilon$  & $M_{scr}^V$ \\
$M_{O}$  &                                           & $\gamma_j\gamma_k$  & $1^{+-}$  & $h_b$        & $M_{scr}^{AV}$ \\ 
\hline
\end{tabular}
\caption{The staggered phases, the $\Gamma$ matrices, the bottomonium states and the corresponding 
screening masses considered in this study.}
\label{tab:meson}
\end{table}
In this study we used point sources corresponding to the quark and antiquarks in the meson correlators and performed  
two state fits of the corresponding correlators to Eq. (\ref{fit}) in order to determine the bottomonium screening 
masses.

\section{Results} 
\begin{figure}
\includegraphics[width=8cm]{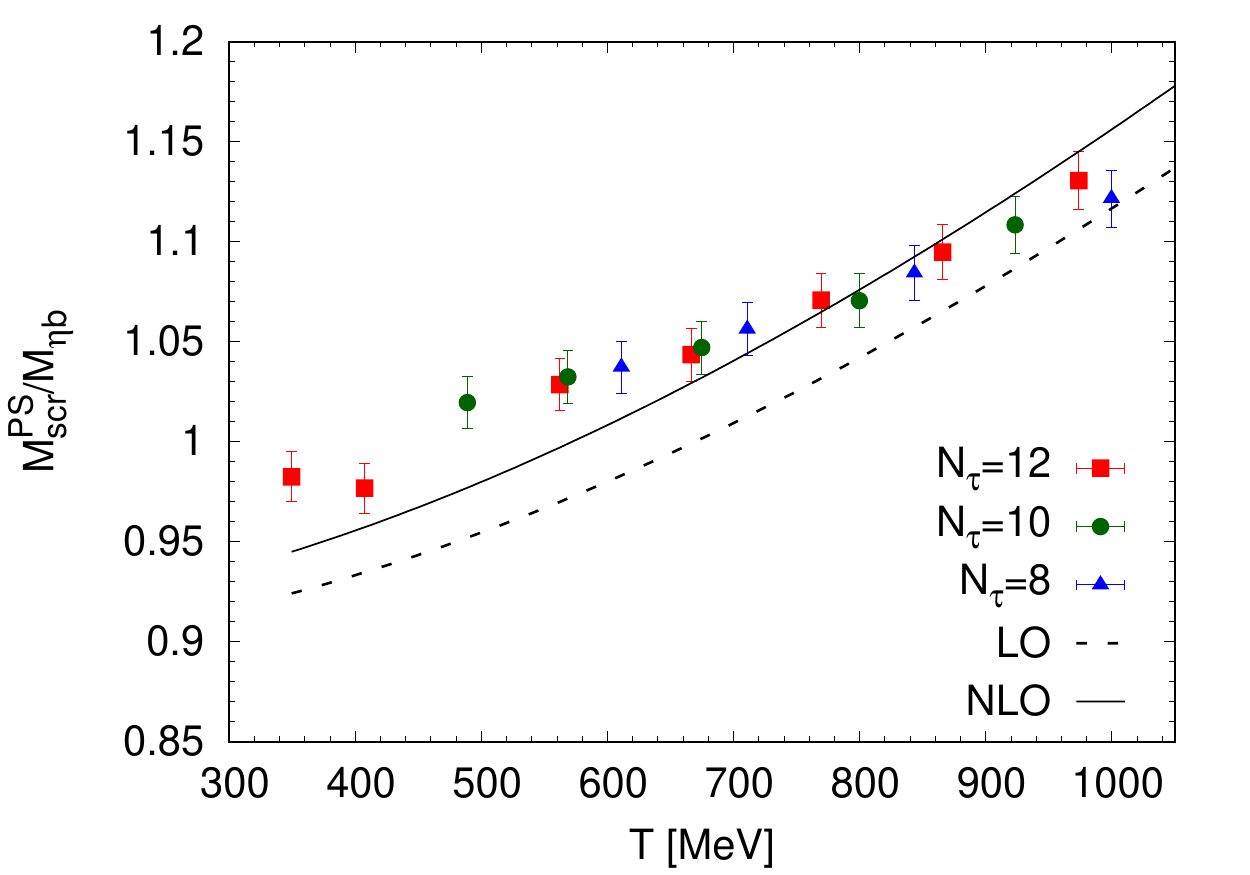}
\caption{The pseudo-scalar screening mass divided by the mass of $\eta_b(1S)$ meson at zero temperature
as a function of the temperature obtained on lattices with $N_{\tau}=8,~10$ and $12$.
The solid line is LO prediction for the screening mass, while the dashed line is the NLO prediction, see text.}
\label{fig:PS}
\end{figure}
\begin{figure}
\includegraphics[width=8cm]{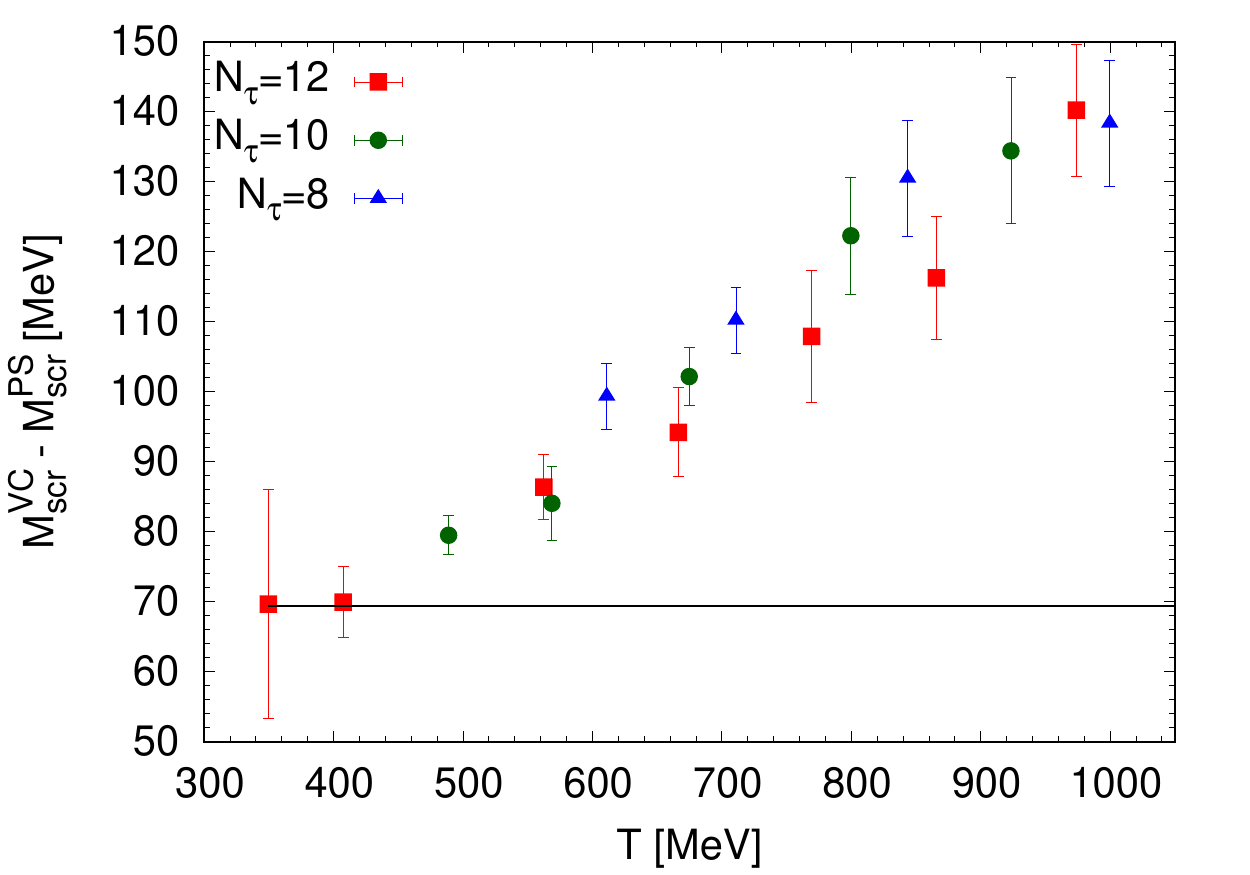}
\caption{The difference between the vector and pseudo-scalar screening masses
as a function of the temperature obtained on lattices with $N_{\tau}=8,~10$ and $12$.
The solid line corresponds to the difference between the $\Upsilon(1S)$ mass and the $\eta_b(1S)$ mass 
from the Particle Data Group (PDG) \cite{PDG20}.
}
\label{fig:VC-PS}
\end{figure} 

\begin{figure}
\includegraphics[width=8cm]{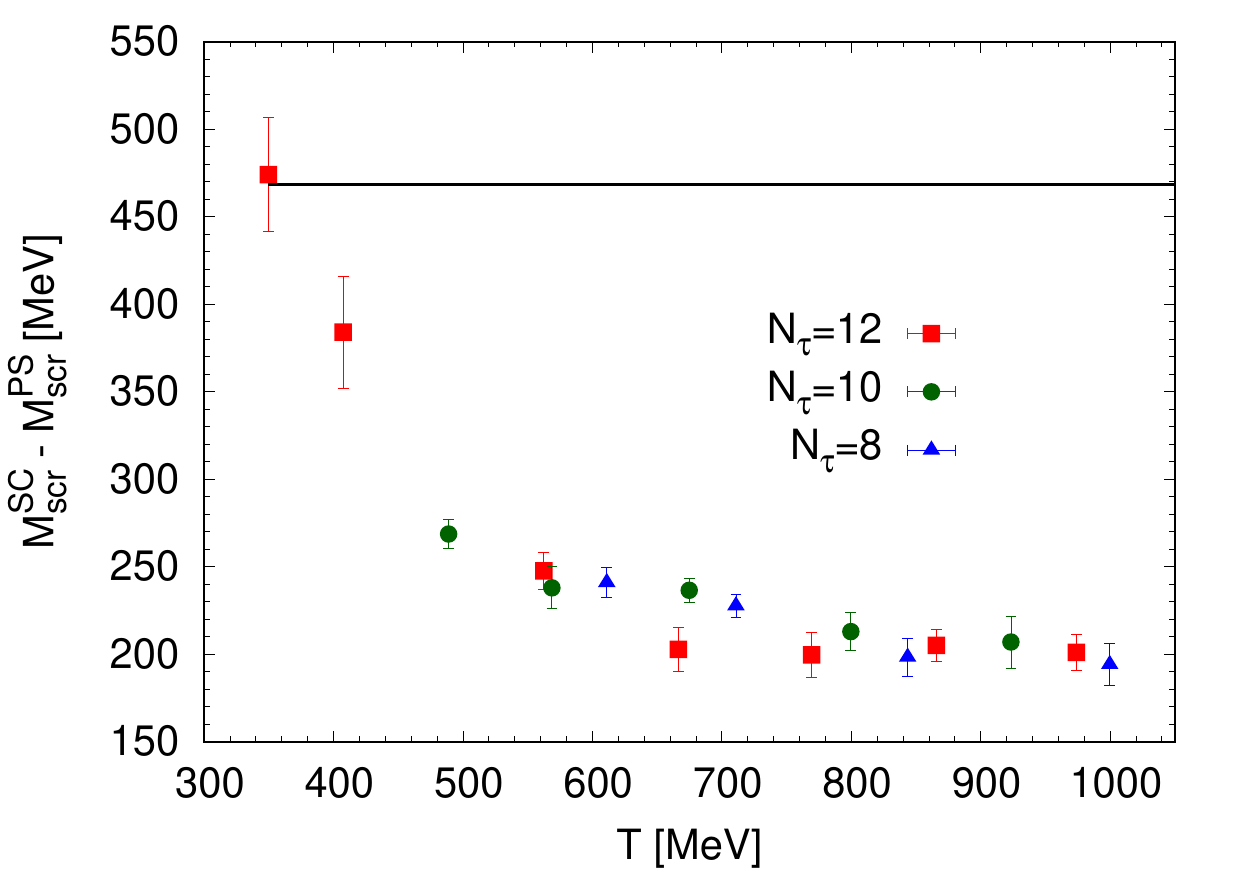}
\caption{The difference between the scalar and pseudo-scalar screening masses
as function of the temperature obtained on lattices with $N_{\tau}=8,~10$ and $12$.
The solid line corresponds to the difference between the $\eta_b(1S)$ and the $\chi_{b0}(1P)$ mass from PDG.
}
\label{fig:SC-PS}
\end{figure}

We begin by showing the pseudo-scalar $\eta_b$ screening mass as a function of the temperature, 
in Fig. \ref{fig:PS}. The results on the screening mass at each temperature have been normalized 
by the zero temperature $\eta_b$ meson mass in this figure. While the ratio $m_b/m_s$ is 
chosen close to its physical value, the lines of constant physics for the strange
quark mass have not been fixed very precisely for this temperature range~\cite{Bazavov:2014pvz}. 
Therefore, we cannot use the experimentally measured mass for $\eta_b$  from the Particle 
Data Group \cite{PDG20}, and need to  take into account the small deviations of
the $b$ quark mass from its physical value. The dependence of the $\eta_b$ meson mass on the $b$ quark mass for the HISQ 
action has been studied earlier in Ref. ~\cite{Petreczky:2019ozv} for $\beta=7.596,~
7.825,~8.0,~8.2$ and $8.4$. Therefore, we could estimate the zero temperature $\eta_b$ mass for 
$\beta=7.825,~8.0,~8.2$ and $8.4$ and the $b$ quark masses given in Table \ref{tab:latpar}.
It turns out that the $\eta_b$ mass is larger than the PDG value by 9.8\% 
for $\beta=8.4$. For $\beta>8.4$, where we do not have the zero 
temperature mass data,  we assume that the $\eta_b$ mass is 9.8\% larger than the experimentally 
measured value based on the above result. 
By fitting the lattice $b$ quark mass that corresponds to 
the physical value for $\beta=7.596,~7.825,~8.0,~8.2$ and $8.4$ \cite{Petreczky:2019ozv}
with the renormalization group inspired Ansatz we determine that the input $b$ quark mass for $\beta=7.65$ is $5.5\%$ larger
than its physical value. We can also estimate that at the close-by value of $\beta$, namely $\beta=7.596$ the $5.5\%$ 
larger $b$-quark mass leads to an $\eta_b$ meson mass that is $4.1\%$ larger than the physical value. 
Therefore, we assume that the $\eta_b$ meson mass for $\beta=7.65$ is $4.4\%$ larger than the PDG value.
Since we did not calculate the 
zero temperature $\eta_b$ masses explicitly on the given lattice spacings but estimated them based on 
the interpolation, we assign a systematic error of $1\%$ to  all zero temperature $\eta_b$ masses
to account for possible systematic effects. For the data in Fig.~ \ref{fig:PS},  we also consider
the errors in the scale setting $a/r_1$ as well as the error of $r_1$ in physical units. Different sources
of errors have been added in quadrature to determine the error on each data point. We observe that  
the lattice cutoff ($N_{\tau}$-dependence) of the results shown in Fig. \ref{fig:PS} is small compared 
to the estimated systematic and statistical errors.

In Fig. \ref{fig:PS} we observe that at the lowest temperatures, the $\eta_b$ screening mass 
is close to the zero temperature mass, while at high temperature the screening mass increases 
roughly linearly with the temperature. The temperature dependence of the $\eta_b$ screening mass 
is qualitatively very similar to the temperature dependence of the ground state charmonium
($\eta_c$ and $J/\psi$) screening masses, except that for charmonium the linear increase with 
the temperature is seen already at $T>250$ MeV \cite{Bazavov:2014cta}. We recall here that the 
approximately linear increase of the screening masses with temperature corresponds to an unbound quark- antiquark 
pair,  where the screening mass at leading order (LO) for quarks of mass $m_q$ is given by 
$2 \sqrt{(\pi T)^2+m_q^2}$. The next-to-leading correction to the screening mass has also been
calculated~\cite{Laine:2003bd}. We show both the LO and NLO result for the bottomonium screening 
mass in Fig.~\ref{fig:PS}.  For the bottom quarks we use the $\overline{MS}$ mass at the scale 
$\mu=m_b$, given by $m(\mu=m_b)=4.188$ GeV \cite{Petreczky:2019ozv}. We observe that the lattice results for the pseudo-scalar 
screening 
mass are close to the NLO predictions for $T>500$ MeV.
The temperature dependence of the pseudo-scalar screening masses for $T>500$ MeV suggests that the bottom quark and 
antiquark are no longer consistently bound, i.e. the $\eta_b$ meson melting is under way at $T>500$ MeV. 
At lower temperatures the $\eta_b$ state exists with small in-medium modifications. The latter conclusion is 
consistent with the findings from NRQCD based studies~\cite{Aarts:2010ek,Aarts:2014cda,Kim:2018yhk,Larsen:2019bwy,Larsen:2019zqv} 
as well as with the results obtained from potential models with a screened complex potential~\cite{Petreczky:2010tk,Burnier:2015tda}.

Next we study the temperature dependence of the difference between the vector $\Upsilon$ and $\eta_b$ 
screening masses, which is shown in Fig. \ref{fig:VC-PS}. We do not expect this difference to be affected by 
the small deviations of the $b$-quark mass from its physical value, and therefore, we do not attempt to correct 
for these small deviations. In estimating the errors for this observable, we have simply added  the errors 
in the determination of the lattice spacings and the statistical errors in quadrature. We again observe a mild
$N_{\tau}$ dependence of the results  compared to the estimated errors. At zero temperature the difference 
between the $\Upsilon(1S)$ and $\eta_b(1S)$  mass is about $70$ MeV \cite{PDG20}, and is caused by spin-dependent interactions, 
which are suppressed as $1/m_b^2$. At the lowest two temperatures the difference between the vector and 
pseudo-scalar screening masses is consistent with this value. This suggests that at these temperatures the 
$\eta_b(1S)$ and $\Upsilon(1S)$ exist as well defined bound states with possibly little in-medium modifications.  
For $T \ge 500$ MeV this difference increases linearly with temperature.  Perturbative calculation at NLO in 
the strong-coupling constant predicts this difference to be identically zero.  In order to understand the 
linear temperature dependence of the difference between the vector and pseudo-scalar screening 
masses at high temperatures one has to go beyond NLO, and instead consider a 
dimensionally reduced three dimensional effective theory of QCD. Within this effective theory, a 
quark and antiquark propagating along the $z$-direction interact via a spin-dependent potential 
which is proportional to the temperature~\cite{Koch:1992nx,Shuryak:1993kg}. This spin dependent 
potential  causes a splitting between the pseudo-scalar and vector screening mass that is also 
proportional to the temperature~\cite{Koch:1992nx,Shuryak:1993kg}. For light quarks, where the
effect of their masses is negligible, this feature has been observed in the lattice calculations for 
$T>900$ MeV, and the difference is $\sim 0.3 T$ \cite{Bazavov:2019www}. For bottom quarks, however, 
the effective quark mass in the effective three dimensional theory is larger,
resulting in the suppression of the spin-dependent interactions. As a result,
the difference between the vector and pseudo-scalar screening masses is smaller than for the light quarks 
in the studied temperature region.  At much higher temperatures, $T\gg m_b$ we expect that this  
difference will eventually approach the value of $0.3T$ even for the bottomonium.
Therefore, the increase in the difference between the vector and pseudo-scalar screening masses
shown in Fig. \ref{fig:VC-PS} is in fact expected and is consistent with an unbound bottom quark anti-quark pair.
This again confirms our assertion that ground state bottomonium melts at $T>500$ MeV.

We now examine the difference between the $\eta_b$ and $\chi_{b0}$, i.e. pseudo-scalar and scalar screening 
masses, as well as the difference between the $\Upsilon$ and $h_b$ masses. Again we do not expect these 
observables to be sensitive to the small error in our determination of the bottom-quark mass. In Fig. ~\ref{fig:SC-PS} 
we show the difference between the scalar and pseudo-scalar screening masses; the error on each data point 
is determined by adding the statistical and scale-setting errors in quadrature. 
As one can see from Fig. ~\ref{fig:SC-PS} the cutoff dependence of the result is mild compared to the estimated errors.
The difference between the axial-vector 
and vector screening masses is very similar to the one shown in the figure, hence not shown explicitly. 
At the lowest temperature $T=350$ MeV,  the difference between the scalar 
and pseudo-scalar (or between the axial-vector and vector) screening masses  agrees with the  differences between 
the $\chi_{b0}(1P)$ and $\eta_b(1S)$ ($h_b(1P)$ and $\Upsilon(1S)$) masses reported in the PDG \cite{PDG20}.
This again suggests that $\chi_{b0}(1P)$ and $h_b(1P)$ states exist in the deconfined medium at 
$T \le 350$ MeV with relatively small medium modifications. This is also consistent with the recent lattice results
using NRQCD which show almost no medium mass shift of the $h_b(1P)$ and $\chi_{b0}(1P)$ states~\cite{Larsen:2019bwy,Larsen:2019zqv}. 
For $T>350$ MeV the difference between the scalar and pseudo-scalar (axial-vector and vector) screening masses decreases with 
increasing temperature, initially very rapidly, by about a factor of two in the temperature 
region $350~{\rm MeV} \le T \le 600~{\rm MeV}$, and then more slowly for $T>600$ MeV. 
At very high temperatures the splitting 
between the scalar and the pseudo-scalar screening masses is expected to be extremely small. In the 
light quark sector, this observation is related to the restoration of chiral and the effective restoration 
of axial U(1) symmetry. For bottom quarks these symmetries are explicitly broken by the large value of 
the quark mass.  However, for $T\gg m_b$ we expect that the difference between the scalar and pseudo-scalar 
screening  correlators will eventually vanish. Therefore, the first rapid drop in the difference between the scalar and
the pseudo-scalar screening mass shown in 
Fig.~\ref{fig:SC-PS} is related to the melting of $\chi_{b0}(1P)$ and $h_b(1P)$ states, while the subsequent 
slower decrease for $T>600$ MeV is related to the tiny effects of the bottom-quark mass and their eventual 
disappearance in the limit of very high temperatures. 

\section{Conclusions}
We performed a first comprehensive study about the temperature dependence of the pseudo-scalar, vector, scalar and axial-vector 
bottomonium screening masses on the lattice using a relativistic (HISQ) action for the bottom quarks.  We scanned a wide range 
of the temperature ranging from $350$ MeV to $1000$ MeV, and for most temperatures performed these calculations at three different 
lattice cutoffs corresponding to temporal extent $N_{\tau}=8,~10$ and $12$ of the lattices. 
We have found the lattice spacing dependence of our results is small compared to other sources
of errors and thus does not effect our main conclusion.
At the lowest temperature, all four screening masses agree with the corresponding bottomonium masses at zero temperature. 
For the axial-vector and scalar screening masses we find a rapid change as a function of temperature for $T>350$ MeV, 
while for vector and pseudo-scalar screening masses the corresponding thermal modifications occurs at a higher temperature, $T>450$ MeV. 
The small thermal modifications of the ground state bottomonium screening masses ($\eta_b(1S)$ and $\Upsilon(1S)$) for $T<450$ MeV are consistent 
with the lattice calculations within the NRQCD formalism~\cite{Aarts:2010ek,Aarts:2014cda,Kim:2018yhk}.
On the other hand we predict that the $1P$ bottomonia will melt at temperatures, somewhere about $350$ MeV, while
the ground state bottomonia will melt at $T>500$ MeV.

\section*{Acknowledgments}
PP was supported by U.S. Department of Energy under Contract No. DE-SC0012704. 
JHW work was supported by U.S. Department of Energy, Office of Science, 
Office of Nuclear Physics and Office of Advanced Scientific Computing Research within 
the framework of Scientific Discovery through Advance Computing (SciDAC) award 
Computing the Properties of Matter with Leadership Computing Resources, and by 
the Deutsche Forschungsgemeinschaft (DFG, German Research Foundation) - Projektnummer 
417533893/GRK2575 ``Rethinking Quantum Field Theory''. SS gratefully acknowledges 
financial support from the Department of Science and Technology, Govt. of India, through a 
Ramanujan Fellowship. 
The numerical calculations have been performed on UQSCD clusters in JLab and FNAL.

\bibliography{ref.bib}

\end{document}